\documentclass[%
 reprint,
 aip,
 jcp,
 numerical,
]{revtex4-1}

\usepackage{gensymb}
\usepackage{hyperref}

\usepackage[version=3]{mhchem}
\usepackage{mathtools}
\usepackage[utf8]{inputenc}
\usepackage{amsfonts,amssymb}
\usepackage{amsmath}
\usepackage{esint}
\usepackage{graphicx}

\begin{document}
\title{Efficiency of a Multi-Reference Coupled Cluster method}

\author{E. Giner}%
\affiliation{Dipartimento di Scienze Chimiche e Farmaceutiche, \\Universita di Ferrara,
Via Fossato di Mortara 17,\\ I-44121 Ferrara, Italy}
\author{G. David}%
\affiliation{Laboratoire de Chimie et Physique Quantiques (CNRS 5626), IRSAMC, Université P. Sabatier, Toulouse (France)}
\author{A. Scemama}%
\email{scemama@irsamc.ups-tlse.fr}
\affiliation{Laboratoire de Chimie et Physique Quantiques (CNRS 5626), IRSAMC, Université P. Sabatier, Toulouse (France)}
\author{J. P. Malrieu}%
\affiliation{Laboratoire de Chimie et Physique Quantiques (CNRS 5626), IRSAMC, Université P. Sabatier, Toulouse (France)}

\begin{abstract}
  The multi-reference Coupled Cluster method first proposed by Meller
  \textit{et al} (\textit{J. Chem. Phys.}
  1996)\cite{Meller_Malrieu_Caballol_1996} has been implemented and tested.
  Guess values of the amplitudes of the single and double excitations (the ${\hat T}$
  operator) on the top of the references are extracted from the knowledge of
  the coefficients of the Multi Reference Singles and Doubles Configuration
  Interaction (MRSDCI) matrix. The multiple parentage problem is solved by
  scaling these amplitudes on the interaction between the references and the
  Singles and Doubles. Then one proceeds to a dressing of the MRSDCI matrix
  under the effect of the Triples and Quadruples, the coefficients of which are
  estimated from the action of ${\hat T}^2$. This dressing follows the logics of the
  intermediate effective Hamiltonian formalism. The dressed MRSDCI matrix is
  diagonalized and the process is iterated to convergence. The method is tested
  on a series of benchmark systems from Complete Active Spaces (CAS) involving 2 or 4
  active electrons up to bond breakings. The comparison with Full Configuration
  Interaction (FCI) results shows that the errors are of the order of a few
  milli-hartree, five times smaller than those of the CASSDCI. The method is
  totally uncontracted, parallelizable, and extremely flexible since it may be
  applied to selected MR and/or selected SDCI. Some potential generalizations
  are briefly discussed.
\end{abstract}

\maketitle

\section{Introduction}

In the domain of molecular physics and quantum chemistry the many-body problem
is perfectly clear as long as it is formulated from a single reference. The
perturbative expansion of the wave operator and its diagrammatic transcription
offer a guide to understand the relations between the multiplicative structure
of the wave function and the additive structure of the energy. The linked
cluster theorem\cite{Goldstone} clarifies the questions of the size consistency
and of the strict separability into fragments. The defects of truncated
Configuration Interaction (CI) are well understood and algorithms have been
proposed to respect approximately
(CEPA-0\cite{Kelly_Sessler_1963,Kelly_Sessler_1964},
CEPA-$n$\cite{Meyer_1971,Meyer_1973,Meyer_1974}), or strictly
((SC)$^2$CI)\cite{Daudey_Heully_Malrieu_1993} the cancellation of unlinked
diagrams. The Coupled Cluster (CC) method\cite{CC_Nucl1,CC_Nucl2,CC_Quantum} is
definitely the most elegant formalism and can be considered as the standard
treatment in its CCSD version, or in the CCSD(T)
version\cite{Barlett_et_al_1990} which incorporates the fourth order effect of
the triply excited determinants. But all these approaches fail when
one cannot expect that a single determinant will represent a reliable starting
point to conveniently generate the wave function.

This is precisely the situation in many domains. The excited states present an
intrinsic multi-determinantal character, and frequently a multi-configurational
character. So are the magnetic systems in their low energy states, and the
treatment of chemical reactions, in which chemical bonds are broken, also
requires to consider geometries for which a single determinant picture is not
relevant. A generalized linked cluster theorem has been established by
Brandow\cite{Brandow_1967}, which gives a conceptual guide, but the conditions
that must be fulfilled for its demonstration (Complete Active Space (CAS) as
reference space, mono-electronic zero-order Hamiltonian) would lead to strongly
divergent behaviors of the corresponding perturbative expansion in any
realistic molecular problem. Practical computational tools have been proposed,
most of them being state-specific. One may quote second order perturbation
expansions based on determinants from selected references
(CIPSI)\cite{Huron_et_al_1973,Evangelisti_et_al_1983}, intermediate
Hamiltonian dressing\cite{Malrieu_Hamiltonien}, in the so-called shifted Bk
technique\cite{Kirtman_1981}. These methods are not strictly size-consistent,
the conditions to satisfy the strict separability of determinant-based
expansions require to define sophisticated zero-order
Hamiltonians\cite{Heully_et_al_1996}. Contracted perturbative expansions, which
perturb the multideterminant zero-order wave function under the effect of
linear combinations of outer-space determinants have also been proposed. One
may quote the CASPT2 method\cite{Andersson_et_al_1990,Andersson_et_al_1992},
which uses a monoelectronic zero-order Hamiltonian, faces intruder state
problems and is not size consistent, the NEVPT2
method\cite{Angeli_et_al_2001_JCP,Angeli_et_al_2001_CPL,Angeli_et_al_2002}
which uses a bi-electronic zero-order Hamiltonian (the Dyall’s
one\cite{Dyall_1995}) and is size consistent and intruder-state free, and the
method from Werner\cite{Celani_et_al_2000}, as well as the perturbation derived
by Mukherjee \textit{et
al}\cite{Ghosh_et_al_2002,Mahapatra_et_al_1999,Sen_et_al_2015} from their MRCC
formalism. 

If the CASSCF wave function is considered as the counterpart of the single
determinant reference the CASSDCI is the counterpart of the SDCI, with the same
size-inconsistence defect, and the research of MRCEPA and MRCC has been the
subject of intense methodological researches for about 20 years, without
evident success. The cancellation of all unlinked terms in the MR expansion
(i.e. a MRCEPA or MR(SC)$^2$CI) formalism is not an easy
task\cite{rev_Szalaay_et_al_2012,rev_Bartlett_2012}. If one lets aside the MRCC
methods that attribute a specific role to a single reference\cite{Adamowicz}, a
few state-specific strictly multi-reference CC methods have been proposed, one
by one of the authors and collaborators\cite{Meller_Malrieu_Caballol_1996},
another one by Mukherjee and coworkers\cite{Mahapatra_et_al_1998}, a third
one in a Brillouin Wigner context\cite{Masik_et_al_1998,Hubac_et_al_2000}. We
return here on the first proposal which had only be tested on a single problem.
We shall present briefly the method in section 2, then the principle of its
implementation (section 3), followed by numerical illustrations of its accuracy
(section 4). The last section will discuss the advantages of this formalism,
its flexibility and possible extensions.  \section{Method}
\subsection{Principle}

Let us call $|I\rangle$ the reference determinants, the number of which will be called $N$.
The reference space may be a CAS,
but this is only compulsory if one wants to satisfy the strict-separability
property. If not the method is applicable to incomplete model spaces as well.
The projector on the model space is
\begin{equation}
  {\hat P}_0=\sum_{I} | I \rangle \langle I |.
\end{equation}
Let us consider a zero-order wave function restricted to the model space,
\begin{equation}
| \Psi_0^m \rangle =\sum_{I} c_I^m | I \rangle.
\end{equation}
This function may be either the eigenfunction of ${\hat P}_0 {\hat H} {\hat P}_0$,
\begin{equation}
  {\hat P}_0 {\hat H} {\hat P}_0| \Psi_0^m \rangle = E_0^m | \Psi_0^m \rangle,
\end{equation}
or the projection of the eigenvector of the CASSDCI on the model space 
\begin{equation}
  | \Psi_0^m \rangle = {\hat P}_0 | \Psi_{\rm CASSDCI}^m \rangle.
\end{equation}
The CASSDCI wave function is written as 
\begin{equation}
  | \Psi_{\rm CASSDCI}^m \rangle = \sum_{I} c_I^m | I \rangle + \sum_{i} c_i^m | i \rangle
\end{equation}
where $|i\rangle$ are the Singles and Doubles (the determinants of the CASSDCI space which do not belong to the reference space).
We want to follow a Jeziorski-Monkhorst\cite{Jeziorski_Monkhorst_1981} expression of the wave operator ${\hat \Omega}$ which is supposed to send from the zero-order wave function to the exact one
\begin{equation}
  {\hat \Omega} | \Psi_0^m \rangle = | \Psi^m \rangle
\end{equation} 
as a sum of reference-dependent operators
\begin{equation}
  {\hat \Omega} {\hat P}_0 = \sum_I {\hat \Omega}_I | I \rangle \langle I | .
\end{equation}
Each of the ${\hat \Omega}_I$’s will take an exponential form
\begin{equation}
  {\hat \Omega}_I = \exp ({\hat T_I}),
\end{equation}
and each operator ${\hat T_I}$ will be truncated to the single and double excitations,
as one does in the CCSD formalism.

\subsection{The multi-parentage problem and the extraction of guess values of the excitation amplitudes from the CASSDCI eigenvector}

One may easily recognize that there exist some degrees of freedom in the
determination of the wave operators. In the single reference CCSD expansion one
searches for the amplitudes of the excitations sending from the reference $\Phi_0$
to the singly and doubly excited determinants. One evaluates the amplitudes of
the Triples and Quadruples as given by the action of ${\hat T}^2$ on $\Phi_0$,
and the eigenequation is projected on each of the Singles and Doubles. If the
number of Singles and Doubles is $n$, one may write a set of $n$ coupled
quadratic equations on the amplitudes.  But it may be more convenient to guess
a first evaluation of these amplitudes from the coefficients of the Singles and
Doubles in the SDCI matrix, which may be done in a unique manner. From these
amplitudes on may obtain a guess of the coefficients of the Triples and
Quadruples and it is convenient (ensuring for instance a better convergence
than solving coupled biquadratic equation) to write the process as an iterative
dressing of the SDCI matrix, in the spirit of Intermediate Effective
Hamiltonian formalism\cite{Malrieu_Hamiltonien}.

In the multireference context one faces a genealogical problem, sometimes
called the multiple-parentage problem. Actually for a state-specific formalism,
one has only one coefficient for each of the singly and doubly excited
determinants $| i \rangle$. In principle one may decide that this determinant
is obtained from each of the references and one would write then
\begin{equation}
c_i^m=\sum_I d_{Ii}^m c_I^m  \label{c_i^m}
\end{equation}
but one must find a criterion to define the $N$ $d_{Ii}$ amplitudes from the
knowledge of a single coefficient. Returning to a perturbative estimate of the
coefficients of the Singles and Doubles starting from $\Psi_0^m$,
the first-order expression of these coefficients \begin{equation}
\label{ci_pert}
c_i^{m(1)}=\frac{\langle \Psi_0^m | {\hat H} | i \rangle}{E_0^m - \langle i | {\hat H} | i \rangle}= \sum_I c_I^m \frac{\langle I | {\hat H} | i \rangle}{E_0^m - \langle i | {\hat H} | i \rangle}
\end{equation}
suggests that the amplitudes of the excitation operators from the references to the Singles and Doubles might satisfy
\begin{equation}
  \frac{d_{Ii}^m}{d_{Ji}^m} = \frac{\langle I | {\hat H} | i \rangle}{\langle J | {\hat H} | i \rangle}.
\end{equation}
This scaling had been proposed in ref.\cite{Meller_Malrieu_Caballol_1996}. This condition may be expressed as 
\begin{equation}
\label{def_dIi}
d_{Ii}^m = \lambda_i^m \langle I | {\hat H} | i \rangle
\end{equation}
where the quantity $\lambda_i^m$ is the inverse of an energy. Re-injecting this expression in Eq.\eqref{c_i^m} leads to
\begin{equation}
  c_i^m = \lambda_i^m \sum_I c_I^m \langle I | {\hat H} | i \rangle,
\end{equation}
which defines $\lambda_i^m$ as
\begin{equation}
\label{lambda}
\lambda_i^m = \frac{c_i^m}{\langle \Psi_0^m | {\hat H} | i \rangle}
\end{equation}
These are the key equations which define guess values of the amplitudes of the excitations leading from the references to the Singles and Doubles. Notice that we only consider amplitudes for the excitations which correspond to physical interactions, and since ${\hat H}$ is at most bi-electronic, one only introduces single- and double-excitation operators. Finally, we can re-express the CASSDCI wave function as 
\begin{equation}
| \Psi_{\rm CASSDCI}^m \rangle =\sum_I c_I^m \left(1 + \sum_i d_{Ii}^m {\hat T}_{Ii} \right) | I \rangle
\end{equation}
where 
\begin{equation}
{\hat T}_{Ii} | I \rangle = | i \rangle.
\end{equation}

\subsection{Coefficients of the Triples and Quadruples}

One may then generate the Triples and Quadruples $| \alpha \rangle$. Among them
only those which interact with the Singles and Doubles (i.e. which are
generated by the action of ${\hat H}$ on the Singles and Doubles and which do not
belong to the CASSDCI space) have to be considered. One may find the references
with which they present either 3 or 4 differences in the occupation numbers of the
molecular orbitals (MOs).
These reference determinants may be called the {\em grand-parents} of $| \alpha
\rangle$. The comparison between $| \alpha \rangle$ and each of its
grand-parents $| I \rangle$ defines the excitation operator from $| I
\rangle$ to $| \alpha \rangle$ as a triple or quadruple excitation
\begin{equation}
{\hat T}_{I\alpha} | I \rangle = | \alpha \rangle
\end{equation}
which may be expressed in second quantization as the product of 4 (or 3) creation operators and 4 (or 3) annihilation operators
\begin{equation}
{\hat T}_{I \alpha} =a_q^\dagger a_p^\dagger a_n^\dagger a_m^\dagger a_e a_f a_g a_h
\end{equation}
The creations run on active and virtual MOs, the annihilations run on active
and inactive occupied MOs but the number of inactive indices among the creation
and/or among the particles must be equal to 3 or 4, otherwise the determinant
would belong to the CASSDCI space. Knowing the operator, it may be factorized
as the product of two complementary double (or single) excitation operators in
all possible manners (each double excitation keeping untouched the $M_s$ value)  
\begin{equation}
{\hat T}_{I \alpha} =\pm {\hat T}_{I k} {\hat T}_{Il} =\pm {\hat T}_{I v} {\hat T}_{I u} = \cdots
\end{equation}
Then we may write the contribution to the coefficient of $| \alpha \rangle$ issued from the reference $| I \rangle$ as 
\begin{equation}
\label{dialpha}
d_{I\alpha}^m = \sum_{(k,l) \in (I \rightarrow \alpha) } \pm d_{Ik}^md_{Il}^m
\end{equation}
where $\{(k,l) \in (I \rightarrow \alpha)\}$ denotes the couples $(k,l)$ for which
${\hat T}_{Ik} {\hat T}_{Il} = \pm {\hat T}_{I \alpha}$. The sign is governed by the permutation
logics. Then one might write the coefficient $c_\alpha^m$ as
\begin{equation}
c_\alpha^m = \sum_I d_{I \alpha}^m c_I^m .
\end{equation}

\subsection{Dressing of the CASSDCI  matrix}

If one considers the eigenequation relative to $\langle i |$
\begin{align}
  \left( \langle i | {\hat H} | i \rangle - E^m \right) c_i^m &+  \sum_I \langle i | {\hat H} | I \rangle c_I^m + \sum_{j \neq i} \langle i | {\hat H} | j \rangle c_j^m \nonumber \\
                                                              &+ \sum_\alpha \langle i | {\hat H} | \alpha \rangle c_\alpha^m = 0 \label{Eq_EV} 
\end{align}
one may decompose the last term 
\begin{align}
  \sum_\alpha \langle i | {\hat H} | \alpha \rangle c_\alpha^m & = \sum_\alpha \langle i | {\hat H} | \alpha \rangle \sum_I d_{I \alpha}^m c_I^m \nonumber \\
                                                               &= \sum_I \left( \sum_\alpha d_{I \alpha }^m \langle i | {\hat H} | \alpha \rangle \right) c_I^m 
\end{align}
Introducing the quantities
\begin{equation}
 \label{diI}
 \langle i | {\hat \Delta}^m | I \rangle = \sum_\alpha d_{I \alpha}^m \langle i | {\hat H} | \alpha \rangle ,
\end{equation}
one may write the eigenequation \eqref{Eq_EV} as
\begin{align}
  \left( \langle i | {\hat H} | i \rangle - E^m \right) c_i^m &+ \sum_I \left( \langle i | {\hat H} | I \rangle + \langle i | {\hat \Delta}^m | I \rangle \right) c_I^m \nonumber \\ 
                                                              &+ \sum_{j \neq i } \langle i | {\hat H} | j \rangle c_j^m = 0 
\end{align}
which suggests to treat the effect of the Triples and Quadruples as a column dressing of the CASSDCI matrix. 
A similar idea has been exploited in the single-reference CCSD context, which may be presented and managed as an iterative dressing 
of the column between the reference and the Singles and Doubles\cite{heff_ccsd}.
The Coupled Cluster dressed CASSDCI Hamiltonian may be written as ${\hat P}_{\rm CASSDCI} \left( {\hat H} + {\hat \Delta}^m \right) {\hat P}_{\rm CASSDCI}$, which is non-Hermitian. Defining the projector on the Singles and Doubles as
\begin{equation}
  {\hat P}_{\rm SD} = {\hat P}_{\rm CASSDCI} - {\hat P}_0,
\end{equation}
\begin{equation}
  {\hat P}_{\rm CASSDCI} {\hat \Delta}^m {\hat P}_{\rm CASSDCI} = {\hat P}_{\rm SD} {\hat \Delta}^m {\hat P}_0
\end{equation}
one may define an equivalent Hermitian dressing ${\hat \Delta}^{m \prime}$ in the case where one considers the Hermitization of the dressed CASSDCI matrix to be desirable,
\begin{equation}
\label{delta_iI_sym}
\langle i | {\hat \Delta}^{m \prime} | I \rangle = \langle I | {\hat \Delta}^{m \prime} | i \rangle = \langle i | {\hat \Delta}^m | I \rangle
\end{equation}
provided that one introduces a diagonal dressing of the CASSDCI matrix 
\begin{equation}
\label{delta_II}
\langle I | {\hat \Delta}^{m \prime} | I \rangle = - \frac{1}{c_I^m} \left( \sum_i \langle I | {\hat \Delta}^{m \prime} | i \rangle c_i^m \right) 
\end{equation}
The diagonalization of the matrices
${\hat P}_{\rm CASSDCI} \left( {\hat H} + {\hat \Delta}^m \right) {\hat P}_{\rm CASSDCI}$ and
${\hat P}_{\rm CASSDCI} \left( {\hat H} + {\hat \Delta}^{m \prime} \right) {\hat P}_{\rm CASSDCI}$
will give the same desired eigenenergy and eigenvector
\begin{equation}
  {\hat P}_{\rm CASSDCI} \left( {\hat H} + {\hat \Delta}^m \right) {\hat P}_{\rm CASSDCI} | \Psi_{\rm CC}^m \rangle = E_{\rm CC}^m{\hat P}_{\rm CASSDCI} | \Psi_{\rm CC}^m \rangle
\end{equation}
\begin{equation}
{\hat P}_{\rm CASSDCI} \left( {\hat H} + {\hat \Delta}^{m \prime} \right){\hat  P}_{\rm CASSDCI} | \Psi_{\rm CC}^m \rangle = E_{\rm CC}^m {\hat P}_{\rm CASSDCI} | \Psi_{\rm CC}^m \rangle .
\end{equation}
Of course the process has to be iterated, the resulting eigenvector defines new coefficients on both the references and the Singles and Doubles, which lead to new amplitudes, new evaluations of the coefficients of the Triples and Quadruples, new dressings. Since the eigenvectors of the dressed matrices are identical, the two formulations, Hermitian or non-Hermitian, converge to the same solution. The converged solutions are the MRCCSD energy and the MRCCSD amplitudes, which define the exponential wave operator.

\section{Implementation}

The proposed algorithm was implemented in the {\em Quantum Package}\cite{QP}, an 
open-source series of programs developed in our laboratory. The bottleneck of this
algorithm is the determinant comparisons needed to determine the excitation
operators and phases during the reconstruction of the genealogy of the $|\alpha
\rangle$'s. This was made possible thanks to a very efficient implementation of
Slater-Condon's rules\cite{SlaterRules}.

\subsection{General structure}
At each iteration step, one first assigns the values of the $\lambda_i^m$ parameters obtained from the eigenvector of the (dressed) CASSDCI matrix 
according to equation \eqref{lambda}.  
From these parameters $\lambda_i^m$ the amplitudes of the single and double excitations are uniquely defined. 
Then one loops on the Singles and Doubles $| i \rangle$. On each of them one reapplies the excitation operators to generate the 
$| \alpha \rangle$’s. Those which belong to the CASSDCI space are eliminated. 
The parents of $| \alpha \rangle$ (that are all the Singles and Doubles $| k \rangle$’s such that $\langle \alpha|{\hat H} |k\rangle \ne 0$) are generated. 
If one of the $| k \rangle$'s has already been considered in the loop on the $| i \rangle$’s ($k<i$) , 
this $| \alpha \rangle$ has been already generated and taken into account and must not be double counted. 
While generating the parents of $| \alpha \rangle$, its interactions with them, $\langle k | {\hat H} | \alpha \rangle$, are stored. 
At this step, the reference grand-parents $| I \rangle$ are identified as having 3 or 4 differences with $| \alpha \rangle$. 
Then, the excitation operator leading from $| I \rangle$ to $| \alpha \rangle$ is expressed in all possible manners 
as products of two complementary single or double excitations. For each couple of complementary excitations, 
the product of the amplitudes is accumulated to compute $d_{I\alpha}^m$ according to equation \eqref{dialpha}. 
Finally, the product of $\langle k | {\hat H} | \alpha \rangle$ 
with $d_{I \alpha}^m$ is accumulated in $\langle k | {\hat \Delta}^m | I \rangle$ for each parent $| k \rangle$ of $| \alpha \rangle$ 
according to equation \eqref{diI}. 
Once the loop on the $| i \rangle$’s is done, all the $| \alpha \rangle$’s have been generated, and the column dressing is completed. 
Then, in order to fit with a symmetric diagonalization technique, one symmetrizes the dressing as mentioned in the preceding section (equations 
\eqref{delta_iI_sym} and \eqref{delta_II}). 
The dressed CASSDCI matrix is diagonalized and the process is repeated up to convergence of the calculated dressed energy. 
From the computational point of view, this process requires the storing of the dressing columns which scales as $N\times (n_{\rm occ}n_{\rm virt})^2$ where $N$ is the number of determinants in the reference, $n_{\rm occ}$ and $n_{\rm virt}$ are respectively the number of occupied an virtual MOs. 
This amount of memory is reasonable, and does not represent a bottleneck for the present applications. 
Regarding the CPU time, the costly part concerns the handling of the $| \alpha \rangle$’s which scales as $(n_{\rm occ}n_{\rm virt})^4$. 
Nevertheless, the process is perfectly parallelizable as all the work done with the $| \alpha \rangle$’s generated from 
$| i \rangle$ does not depend on the other $| i \rangle$'s.  

\subsection{Practical issues}
The definition of $\lambda_i^m$ can lead to numerical instabilities when 
$\langle \Psi_0^m | {\hat H} | i \rangle$ is small. Nevertheless, in such cases the contribution of $| i \rangle$ 
to the post-CAS correlation energy is also small, suggesting that one might use a perturbative estimate of  $\lambda_i^m$. 
In practice, we use the perturbative $\lambda_i$ according to two different criteria. The first one concerns the ratio of the 
variational coefficient $c_i^m$ (obtained at a given iteration) over its perturbative estimate (see \eqref{ci_pert}). 
If $\frac{c_i^{m(1)}}{c_i^m}\notin [0,0.5]$ then the amplitudes involving $|i\rangle$ are determined using the perturbative $\lambda_i^{m({\rm pert})}$
defined as 
\begin{equation}
\label{lambda_pert}
\lambda_i^{m({\rm pert})} = \frac{1}{E_0^m - \langle i | {\hat H} | i \rangle}
\end{equation}
In such situations the coefficient $c_i^m$ is not determined by its interaction with the reference determinants, but comes from higher-order effects. 
The second criterion concerns the  absolute value of each of the $d_{Ii}^m$ defined according to \eqref{def_dIi}.  
If any of these terms calculated with the $\lambda_i^m$ obtained from the variational calculation (see \eqref{lambda}) is larger than 0.5, 
the perturbative $\lambda_i^{m({\rm pert})}$ is used to determine the amplitudes $d_{Ii}^{m({\rm pert})}$ defined as 
\begin{equation}
\label{dIi_pert}
d_{Ii}^{m({\rm pert})} = H_{Ii} \lambda_i^{m({\rm pert})}
\end{equation}
and the working amplitudes $d_{Ii}^{m}$ are set to $d_{Ii}^{m({\rm pert})}$. This condition avoids numerical instabilities occurring when both 
$c_i^m$ and $\langle \psi_0^m|{\hat H}|i\rangle$ are small, and allows us the control of the maximum value of the amplitudes. 
As soon as along the iterations one of the $|i\rangle$'s fulfills one of these criteria, it will be treated perturbatively in the following 
iterations. This precaution avoids significant oscillations due to back and forth movements from perturbative to variational treatment of the 
$\lambda_i^m$. The numerically observed residual oscillations are of the order of magnitude of $10^{-6}$~$E_{\rm h}$, which may certainly be attributed to 
the non linear character of the numerical algorithm. Nevertheless, the order of magnitude of the residual oscillations is much smaller than 
the chemical and even spectroscopic accuracy. 

\section{Numerical test studies}

We decided to test the accuracy and robustness of the method on a series of
benchmarks, some of which have been used in the evaluations of other MRCC
proposals and of alternative MR approaches. They essentially concern model
problems, especially bond breaking problems or the treatment of degenerate
situations. They require to use a CAS with either two electrons in two MOs or
four electrons in four MOs.  In all cases the method converged in a few
iterations. 
A systematic comparison is made with FCI estimates, either taken from the
literature or obtained from a CIPSI type variation+perturbation
calculation\cite{Huron_et_al_1973,Evangelisti_et_al_1983} where the
perturbative residue is about -6~m$E_{\rm h}$.
Of course the CASSDCI is already a rather sophisticated treatment, which takes
into account, although in a size-inconsistent manner, the leading correlation
effects, both the non-dynamical part in the CAS and the dynamical part in the
SDCI step. One may expect that the improvement brought by the MRCC treatment
will be significant when the number of important inactive double excitations is
large. 

In order to have a global view of the performance of the here-proposed
algorithm, we report for each calculation (except the symmetric
dissociation of the water molecule) potential energy curves, the
error to FCI estimate of our MRCCSD algorithm together with the CASSDCI.  Tables
showing the error with respect to the FCI estimate of the MRCCSD and CASSDCI are
also presented, complemented by the total energies of the FCI estimate. The non-parallelism error (NPE) is here calculated as the difference between the
minimum and maximum error to the FCI estimate.  The spectroscopic constants are
obtained from an accurate fit of the obtained potential energy curves with a
generalized Morse potential representation.  The spectroscopic constants
reported here are the equilibrium distance $R_{\rm eq}$ in \AA, the
frequency $k_{\rm eq}$ in $E_{\rm h}/{\rm \AA}^2$ and the atomization energy $D_e$
in kcal/mol.

All the calculations were performed with the Quantum Package\cite{QP}, an
open-source series of programs developed in our group.

\subsection{Single-bond breakings} 

The treatment of the breaking of a single bond in principle requires only a
CASSSCF zero-order treatment including two electrons in two MOs. We have
considered three problems of that type. 

\subsubsection*{Bond breaking of the F$_2$ molecule}
The F$_2$ molecule is a paradigmatic molecule since it is a case where the dynamical correlation brings a crucial contribution to the bonding. 
Despite the closed shell character of the wave function in the equilibrium region the single reference Hartree-Fock (HF) solution is unbound (by 18~kcal/mol) 
with respect to the restricted open shell HF solution of the fluorine atoms. The 2-electron in 2-MO CASSCF treatment binds the molecule by 18~kcal/mol, 
but the experimental binding energy is much larger (39~kcal/mol). Going to a full valence CASSCF (14 electrons in 10 MOs) does not bring any 
improvement. The role of the dynamical correlation has been extensively studied and may be seen as a dynamic response of the lone pair electrons 
to the fluctuation of the electric field created by the two electrons of the $\sigma$ bond\cite{hiberty_f2,genesis}. 
The concept of orbital breathing has been 
proposed to express the fact that the orbitals of the lone pairs tend to become more diffuse on the negative center and more contracted 
on the positive center in the ionic valence-bond (VB) components of the CAS. These dynamic relaxation processes can only take place if one uses non-minimal 
basis sets.

\begin{figure}[htb!]
\includegraphics[width=\columnwidth]{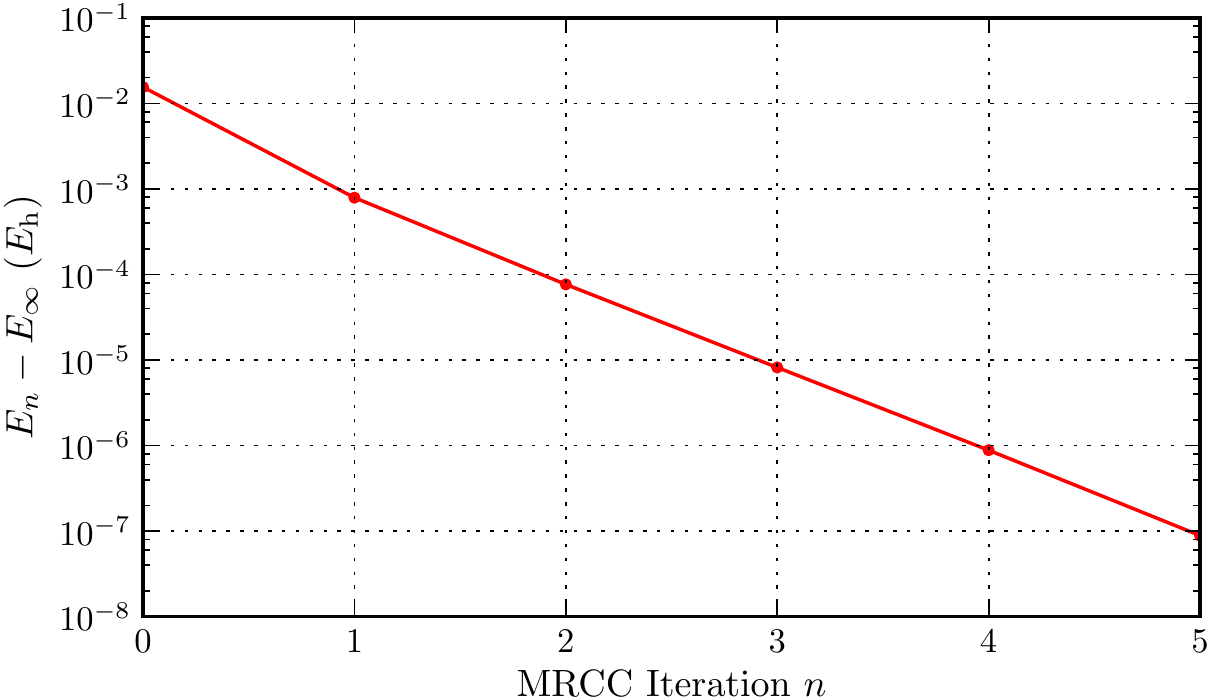}
\caption{F$_2$ molecule at $R$=1.45~\AA, cc-pVDZ basis set. Convergence of the MRCCSD energy along the iterations. }
\label{f2_conv}
\end{figure}

\begin{figure}[htb!]
\includegraphics[width=\columnwidth]{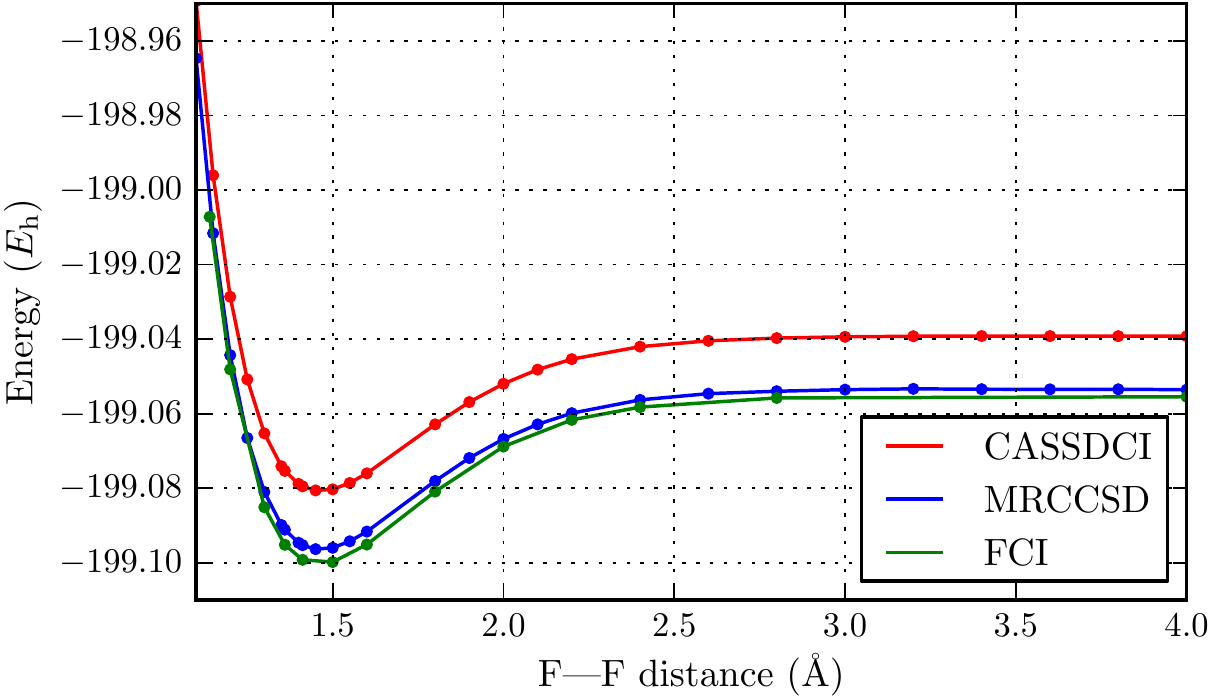}
\caption{Potential energy curves of the F$_2$ molecule, cc-pVDZ basis set. }
\label{f2_pot}
\end{figure}

\begin{figure}[htb!]
\includegraphics[width=\columnwidth]{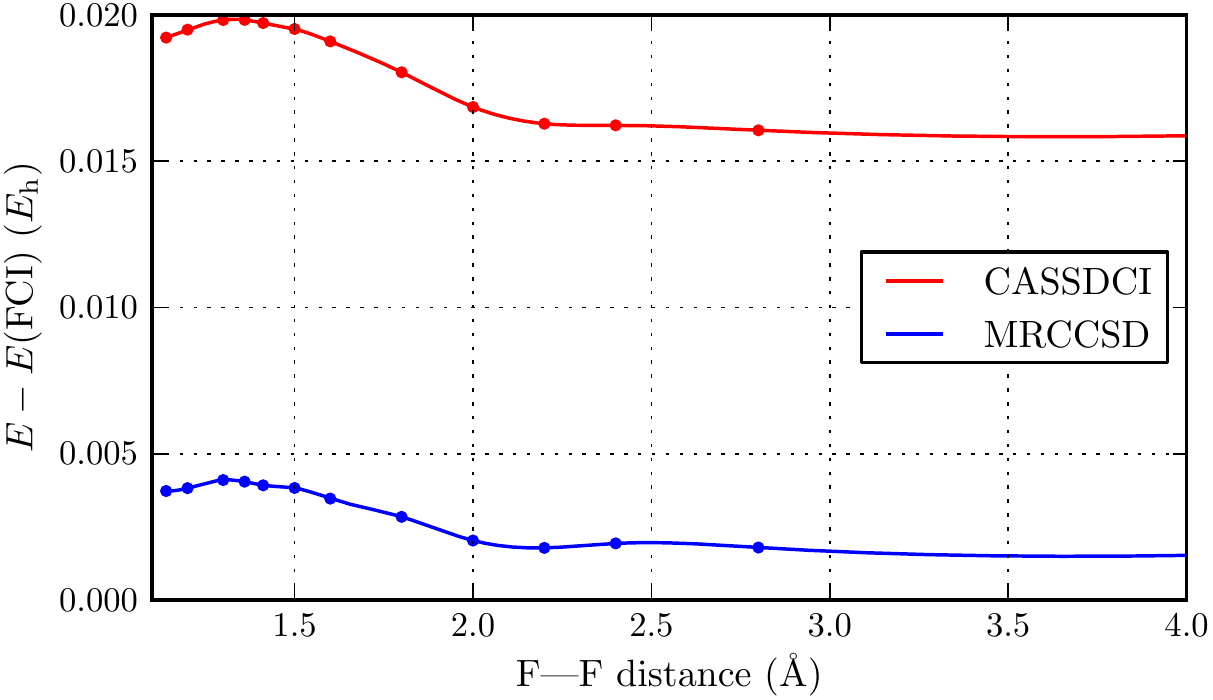}
\caption{F$_2$ molecule, cc-pVDZ basis set. Errors with respect to the FCI estimate as a function of the F---F distance.}
\label{f2_delta}
\end{figure}

The calculations of F$_2$ were obtained in the cc-pVDZ basis set\cite{Dunning_1989} keeping the $1s$ electrons frozen, 
and accurate FCI estimates are taken from the work of by Bytautas \textit{et al}\cite{bytautas}. 
Figure~\ref{f2_conv} shows an exponential convergence of the energy along the MRCC iterations. 
The here-reported calculation, performed in a medium size basis set, does not afford a sufficient flexibility 
to reach the experimental binding energy (the estimated FCI binding energy in this basis is $D_e$=28.3~kcal/mol). 
The potential energy curves and the error to FCI estimate are reported, respectively in Figure~\ref{f2_pot} and Figure~\ref{f2_delta}, and 
the estimated FCI values together with the error of the MRCCSD and CASSDCI calculations appear in Table~\ref{tabf2}. 
The average error is reduced by a factor close to 6, and the NPE is only reduced by 40\%  by the MRCCSD calculations. 

\begin{table}[htb!]
 \begin{center}
\begin{tabular}{lccc}
\hline
\hline
\scriptsize $R$ (\AA)
              & \scriptsize $E_{\rm CASSDCI} - E_{\rm FCI}$  & \scriptsize $E_{\rm MRCCSD} - E_{\rm FCI}$ & \scriptsize FCI estimate \\ 
\hline
     1.14     &                    19.223                    &                   3.726                    &       -199.007 18        \\ 
     1.20     &                    19.495                    &                   3.823                    &       -199.048 11        \\ 
     1.30     &                    19.825                    &                   4.102                    &       -199.085 10        \\ 
     1.36     &                    19.829                    &                   4.045                    &       -199.095 17        \\ 
   1.41193    &                    19.721                    &                   3.920                    &       -199.099 20        \\ 
     1.50     &                    19.518                    &                   3.830                    &       -199.099 81        \\ 
     1.60     &                    19.094                    &                   3.466                    &       -199.095 10        \\ 
     1.80     &                    18.038                    &                   2.843                    &       -199.080 90        \\ 
     2.0      &                    16.850                    &                   2.034                    &       -199.068 82        \\ 
     2.2      &                    16.280                    &                   1.783                    &       -199.061 65        \\ 
     2.40     &                    16.225                    &                   1.936                    &       -199.058 23        \\ 
     2.80     &                    16.055                    &                   1.794                    &       -199.055 77        \\ 
     8.00     &                    16.241                    &                   1.893                    &       -199.055 45        \\ 
\hline                                                                                                                               
\hline
 $R_{\rm eq}$ &                    1.466                     &                   1.465                    &          1.460           \\ 
 $k_{\rm eq}$ &                    0.730                     &                   0.739                    &          0.795           \\ 
   $D_{e}$    &                    26.01                     &                   26.91                    &          28.31           \\ 
\hline
\hline
\end{tabular}
 \end{center}
 \caption{F$_2$ molecule, cc-pVDZ basis set. Total energies are given in
 $E_{\rm h}$, and the energy differences are given in m$E_{\rm h}$.}
 \label{tabf2}
\end{table}

\subsubsection*{The C---C bond breaking in ethane}

\begin{figure}[htb!]
\includegraphics[width=\columnwidth]{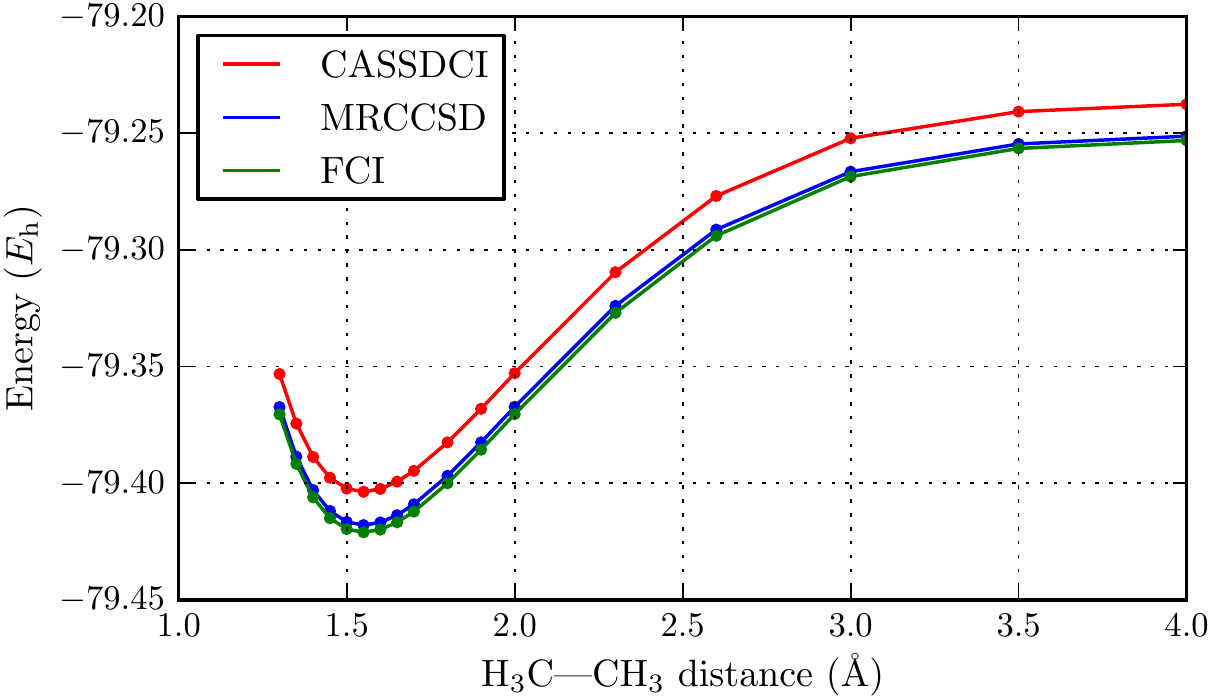}
\caption{Ethane molecule, 6-31G basis set. Potential energy curves along the C---C bond stretching.}
\label{c2h6_pot}
\end{figure}

\begin{figure}[htb!]
\includegraphics[width=\columnwidth]{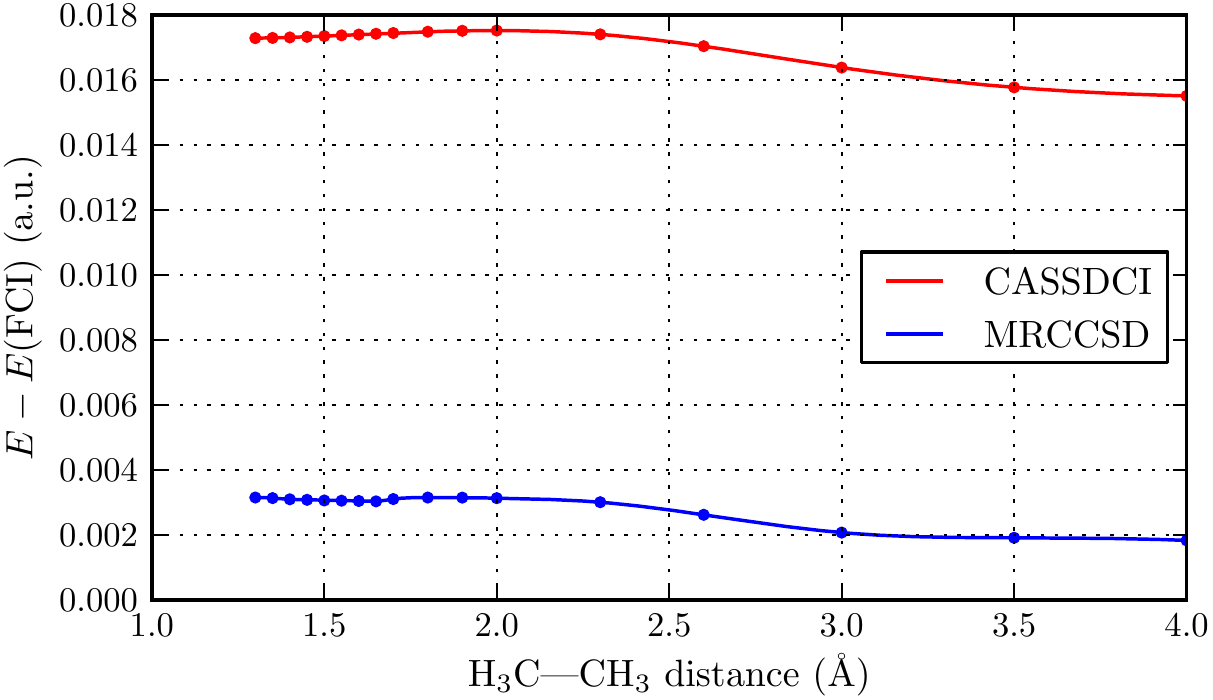}
\caption{Ethane molecule, 6-31G basis set. Errors with respect to the FCI estimate as a function of the C---C distance. }
\label{c2h6_delta}
\end{figure}

This calculation is performed in the 6-31G basis set, keeping the $1s$
electrons frozen.  The geometrical parameters are given in
Table~\ref{tabgeomc2h6}.  The potential energy curves are reported in
Figure~\ref{c2h6_pot} and the errors with respect to the FCI estimate appear in
Figure~\ref{c2h6_delta}.  These data show that the error with respect to the FCI
energy is greatly reduced by a factor of 6 in average. According to
Table~\ref{tabc2h6}, the NPE goes from 2.01~m$E_{\rm h}$ to 1.32~m$E_{\rm h}$
for respectively the CASSDCI and MRCCSD approaches.  Concerning the
spectroscopic constants, the impact of the CC treatment is modest
but goes in the right direction.

\begin{table}[htb!]
 \begin{center}
 \begin{tabular}{lcc}
 \hline
 \hline
 \scriptsize
 Geometrical parameters & C$_2$H$_6$ & C$_2$H$_4$ \\
 \hline
 C---H                          & 1.103~\AA       & 1.089~\AA     \\
 C---C                          & 1.550~\AA       & 1.335~\AA     \\
 H---C---C                      & 111.2\degree & 120.0\degree \\
 H---C---H                      & 107.6\degree & 120.0\degree \\
 H---C---C---H                  & 180.0\degree & 180.0\degree \\
 \hline
 \hline
 \end{tabular}
 \end{center}
 \caption{Geometries used for ethane and ethylene. }
 \label{tabgeomc2h6}
\end{table}

\begin{table}[htb!]
 \begin{center}
 \begin{tabular}{lccc}
 \hline
 \hline
\scriptsize
$R$ (\AA)
            & \scriptsize $E_{\rm CASSDCI} - E_{\rm FCI}$  & \scriptsize $E_{\rm MRCCSD} - E_{\rm FCI}$ & \scriptsize FCI estimate \\ 
\hline                                                                                                                         
   4.00     &                  15.508                  &                   1.834                    &       -79.253 166        \\ 
   3.50     &                  15.770                  &                   1.915                    &       -79.256 574        \\ 
   3.00     &                  16.379                  &                   2.074                    &       -79.268 617        \\ 
   2.60     &                  17.037                  &                   2.622                    &       -79.293 972        \\ 
   2.30     &                  17.402                  &                   3.009                    &       -79.326 999        \\ 
   2.00     &                  17.519                  &                   3.134                    &       -79.370 376        \\ 
   1.90     &                  17.510                  &                   3.150                    &       -79.385 598        \\ 
   1.80     &                  17.482                  &                   3.152                    &       -79.399 969        \\ 
   1.70     &                  17.442                  &                   3.106                    &       -79.412 107        \\ 
   1.65     &                  17.419                  &                   3.035                    &       -79.416 695        \\ 
   1.60     &                  17.395                  &                   3.046                    &       -79.419 813        \\ 
   1.55     &                  17.371                  &                   3.055                    &       -79.420 987        \\ 
   1.50     &                  17.347                  &                   3.062                    &       -79.419 613        \\ 
   1.45     &                  17.326                  &                   3.083                    &       -79.414 941        \\ 
   1.40     &                  17.306                  &                   3.099                    &       -79.406 030        \\ 
   1.35     &                  17.291                  &                   3.135                    &       -79.391 701        \\ 
   1.30     &                  17.284                  &                   3.153                    &       -79.370 480        \\ 
\hline
\hline
 $R_{eq}$     &                  1.549                   &                   1.550                    &          1.550           \\ 
 $k_{\rm eq}$ &                  1.018                   &                   1.017                    &          1.015           \\ 
 $D_{\rm e}$  &                  104.52                  &                   104.99                   &          105.75          \\ 
\hline
\hline
 
 \end{tabular}
 \end{center}
 \caption{Ethane molecule, 6-31G basis set. The FCI estimate is the CIPSI calculation.
   Total energies are given in $E_{\rm h}$, and the energy differences are given in m$E_{\rm h}$.}
 \label{tabc2h6}
\end{table}

\subsubsection*{The rotation of the ethylene molecule around its C---C bond}

\begin{figure}[htb!]
\includegraphics[width=\columnwidth]{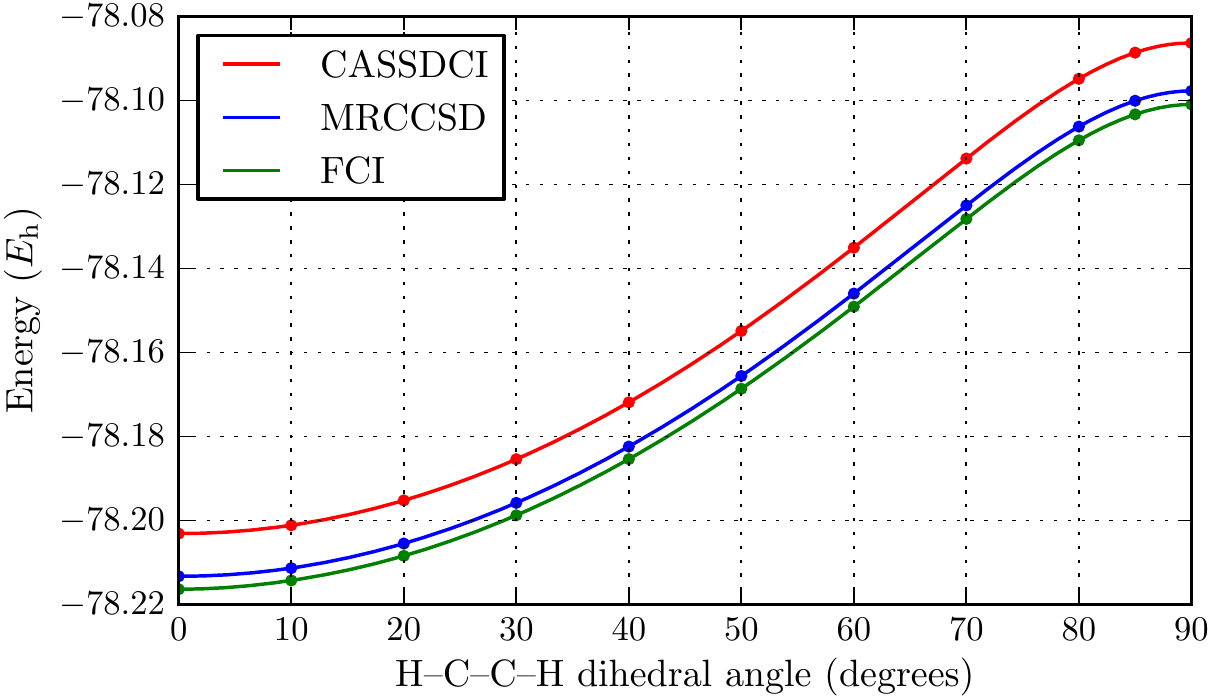}
\caption{Ethylene molecule, 6-31G basis set. Energy as a function of the rotation around the C---C bond.}
\label{c2h4_rot}
\end{figure}

\begin{figure}[htb!]
\includegraphics[width=\columnwidth]{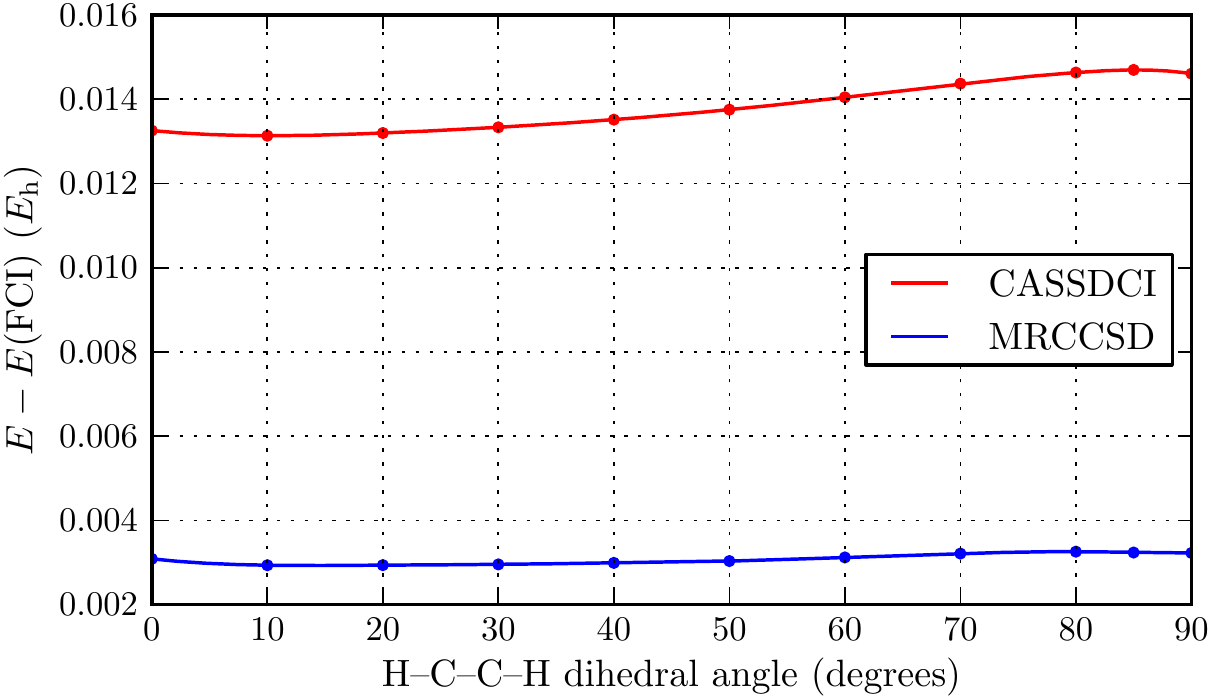}
\caption{Ethylene molecule, 6-31G basis set. Errors with respect to the FCI estimate as a function of the rotation around the C---C bond.}
\label{c2h4_rot_delta}
\end{figure}

This twisting breaks the $\pi$ bond. The calculation is performed in the 6-31G
basis set at the geometry given in Table~\ref{tabgeomc2h6}, keeping the $1s$
electrons frozen.  
The occupied MOs in the inactive space involve 10 electrons, and despite the
modest size of the basis set one may expect a significant size-consistence
defect of the CASSDCI results, since they miss the repeatability of inactive
double excitations on the SD determinants.  The potential energy curve along
the angle of rotation is reported in Figure~\ref{c2h4_rot} and the error to the
FCI estimate is reported in Figure~\ref{c2h4_rot_delta}.  From these data it
appears that the global shape of the potential energy curve obtained using the
CC treatment is more parallel to the FCI curve than using the CASSDCI approach.
From Table~\ref{tabc2h4_rot}, one observes that the error with respect to the
FCI estimate is reduced by a factor of 6 when going from CASSDCI to MRCCSD.  Also,
the NPE is also reduced from 1.6~m$E_{\rm h}$ to 0.3~m$E_{\rm h}$.

\begin{table}[htb!]
 \begin{center}
 \begin{tabular}{lccc}
\hline
\hline
\scriptsize
Angles (degrees)
    & \scriptsize $E_{\rm CASSDCI} - E_{\rm FCI}$ & \scriptsize $E_{\rm MRCCSD} - E_{\rm FCI}$ & \scriptsize FCI estimate  \\ 
\hline                                                                                         
 0  &                    13.255                   &                   2.935                    &       -78.216 340         \\
 10 &                    13.132                   &                   2.935                    &       -78.214 241         \\ 
 20 &                    13.196                   &                   2.938                    &       -78.208 391         \\ 
 30 &                    13.331                   &                   2.955                    &       -78.198 732         \\ 
 40 &                    13.513                   &                   2.991                    &       -78.185 373         \\ 
 50 &                    13.750                   &                   3.035                    &       -78.168 619         \\ 
 60 &                    14.043                   &                   3.120                    &       -78.149 094         \\ 
 70 &                    14.368                   &                   3.212                    &       -78.128 205         \\ 
 80 &                    14.631                   &                   3.258                    &       -78.109 498         \\ 
 85 &                    14.694                   &                   3.237                    &       -78.103 326         \\ 
 90 &                    14.605                   &                   3.227                    &       -78.100 966         \\ 
\hline
\hline
 \end{tabular}
 \end{center}
 \caption{Rotation of the ethylene molecule, 6-31G basis set. The FCI estimate is the CIPSI calculation.
   Total energies are given in $E_{\rm h}$, and the energy differences are given in m$E_{\rm h}$.}
 \label{tabc2h4_rot}
\end{table}

\subsection{Two-bond breakings}

Three systems have been treated using a CAS with four electrons in 4 active
MOs. Two of them concern the simultaneous breaking of two bonds.

\subsubsection*{Breaking of the C=C double bond of ethylene}

\begin{figure}[htb!]
\includegraphics[width=\columnwidth]{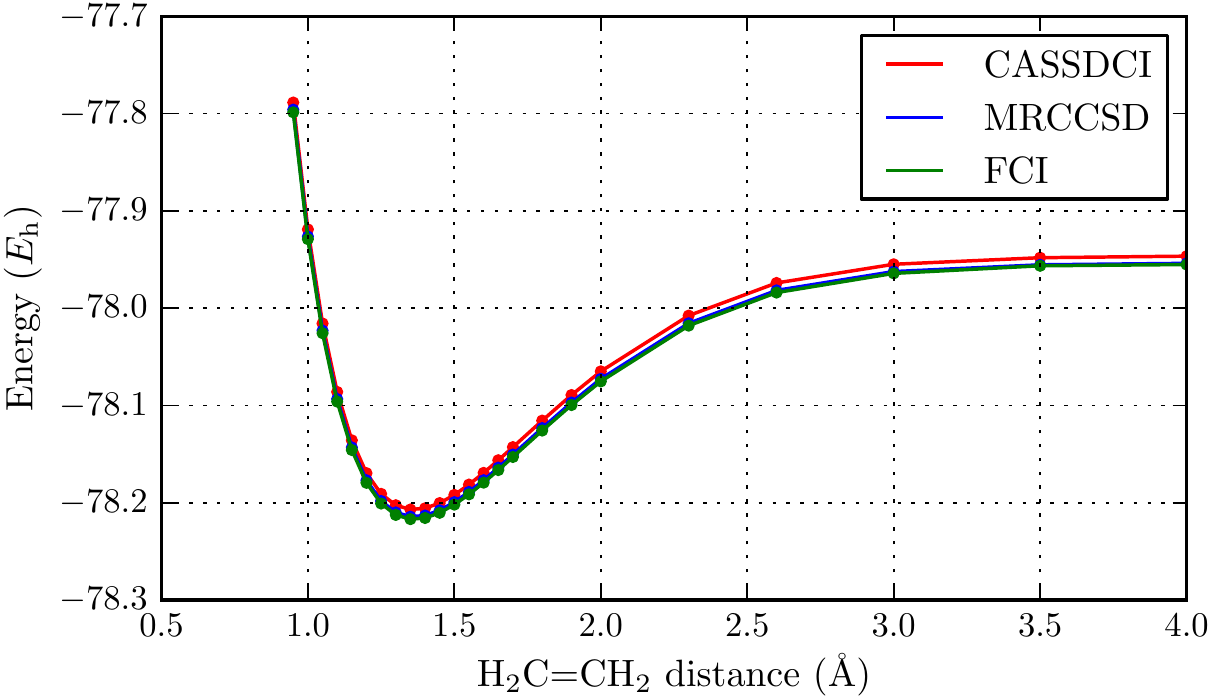}
\caption{Potential energy curves of the ethylene molecule, 6-31G basis set.}
\label{ethylene_pot}
\end{figure}

\begin{figure}[htb!]
\includegraphics[width=\columnwidth]{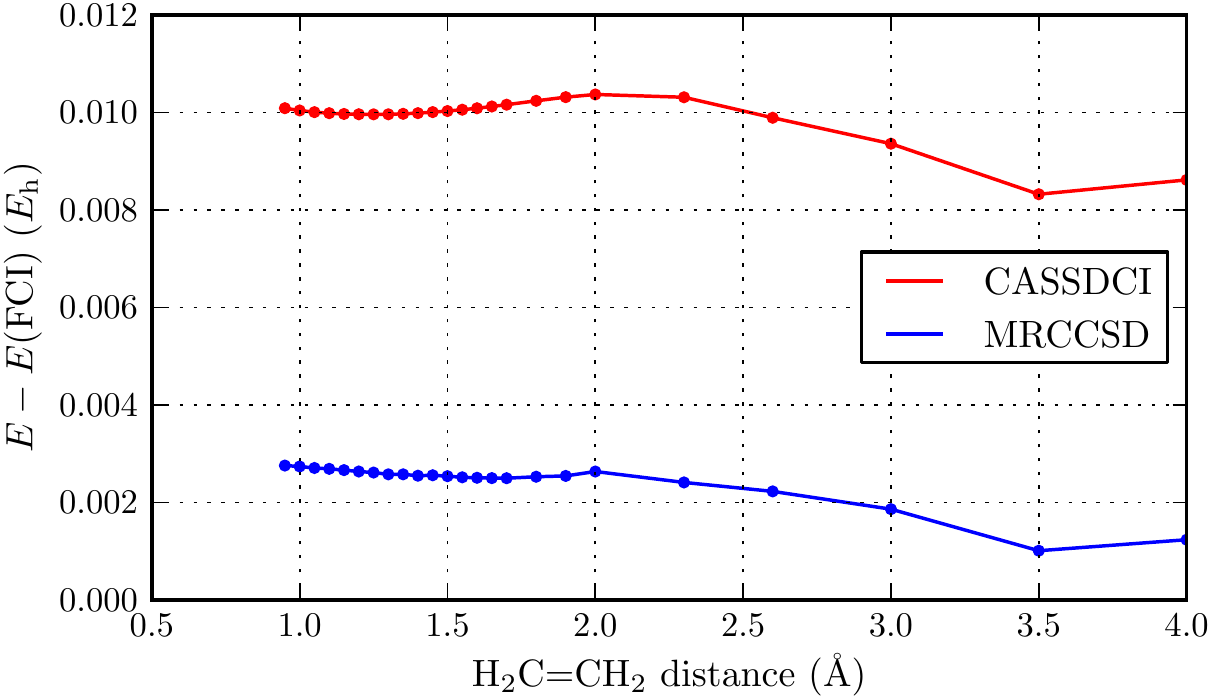}
\caption{Potential energy curves of the ethylene molecule, 6-31G basis set. Errors with respect to the FCI estimate as a function of the C---C distance. }
\label{ethylene_delta}
\end{figure}

The dissociation of the ethylene molecule by breaking the double bond was
studied in the 6-31G basis set, with the geometry given in
Table~\ref{tabgeomc2h6}.  We report the potential energy curves in
Figure~\ref{ethylene_pot}, and the error with respect to the FCI estimate in
Figure~\ref{ethylene_delta}. 
The corresponding values appear in Table~\ref{tabethylene}. Again, the error
to estimated FCI energy is reduced by a factor of 4, but the NPE is reduced
only by 20\% with the CC treatment.

\begin{table}[htb!]
\begin{center}
\begin{tabular}{lccc}
\hline
\hline
\scriptsize
$R$ (\AA)     & \scriptsize $E_{\rm CASSDCI} - E_{\rm FCI}$  & \scriptsize $E_{\rm MRCCSD} - E_{\rm FCI}$ & \scriptsize FCI estimate \\ 
\hline                                                                                                                               
   1.20       &                    9.962                     &                   2.635                    &       -78.179 508        \\ 
   1.25       &                    9.960                     &                   2.614                    &       -78.200 805        \\ 
   1.30       &                    9.961                     &                   2.578                    &       -78.212 507        \\ 
   1.35       &                    9.971                     &                   2.578                    &       -78.216 869        \\ 
   1.40       &                    9.985                     &                   2.549                    &       -78.215 666        \\ 
   1.45       &                    10.006                    &                   2.558                    &       -78.210 313        \\ 
   1.50       &                    10.028                    &                   2.539                    &       -78.201 918        \\ 
   1.55       &                    10.055                    &                   2.517                    &       -78.191 365        \\ 
   1.60       &                    10.085                    &                   2.507                    &       -78.179 345        \\ 
   1.65       &                    10.121                    &                   2.500                    &       -78.166 404        \\ 
   1.70       &                    10.158                    &                   2.498                    &       -78.152 957        \\ 
   1.80       &                    10.238                    &                   2.528                    &       -78.125 775        \\ 
   1.90       &                    10.313                    &                   2.545                    &       -78.099 562        \\ 
   2.00       &                    10.368                    &                   2.635                    &       -78.075 294        \\ 
   2.30       &                    10.310                    &                   2.411                    &       -78.017 944        \\ 
   2.60       &                    9.890                     &                   2.228                    &       -77.983 952        \\ 
   3.00       &                    9.358                     &                   1.863                    &       -77.964 220        \\ 
   3.50       &                    8.321                     &                   1.011                    &       -77.956 421        \\ 
   4.00       &                    8.616                     &                   1.237                    &       -77.955 127        \\ 
\hline
\hline
 $R_{\rm eq}$ &                    1.362                     &                   1.362                    &          1.362           \\ 
 $k_{\rm eq}$ &                    2.043                     &                   2.039                    &          2.042           \\ 
 $D_{e}$      &                    163.48                    &                   163.54                   &          164.47          \\ 
\hline
\hline

\end{tabular}
 \end{center}
 \caption{Dissociation of the ethylene molecule, 6-31G basis set. The FCI estimate is the CIPSI calculation.
   Total energies are given in $E_{\rm h}$, and the energy differences are given in m$E_{\rm h}$.}
 \label{tabethylene}
\end{table}

\subsubsection*{Two-bond breaking in H$_2$O}
This is a rather well known test problem for MRCC methods. The calculation is
done with the cc-pVDZ basis set at five different geometries obtained from the
equilibrium geometry ($R_e$=1.84345~\AA, and $\angle_{\text {HOH}} = 110.6\degree$), 
in order to compare with the values of the literature.\cite{H2O}
The results appear in Table~\ref{tabh2o}. 
The benefit of the MRCCSD with respect to the CASSDCI treatment is significant : 
the maximum error is 1.4~m$E_{\rm h}$, better than the 6.4~m$E_{\rm h}$ given by the Mk-MRCC treatment.
This improvement may be due to the here-proposed treatment of the amplitudes responsible for potential divergences. 
The NPE goes from 2~m$E_{\rm h}$ to 0.7~m$E_{\rm h}$ when the CC treatment is applied. 

\begin{table}[htb!]

\begin{center}
 \begin{tabular}{lcccc}
 \hline
 \hline
 \scriptsize
 $R$ (\AA)   & \scriptsize $E_{\rm CASSDCI} - E_{\rm FCI}$ & \scriptsize $E_{\rm Mk} - E_{\rm FCI}$  & \scriptsize $E_{\rm MRCCSD} - E_{\rm FCI}$ & \scriptsize FCI \\ 
\hline                                                                                                                                                          
 1    $R_e $ &                    4.923                    &               2.909                 &                   1.407                    &    -76.241860   \\ 
 1.5  $R_e $ &                    4.674                    &               4.817                 &                   1.248                    &    -76.072348   \\ 
 2.0  $R_e $ &                    3.665                    &               6.485                 &                   0.855                    &    -75.951665   \\ 
 2.5  $R_e $ &                    3.097                    &               5.672                 &                   0.763                    &    -75.917991   \\ 
 3.0  $R_e $ &                    2.959                    &               3.987                 &                   0.845                    &    -75.911946   \\ 
 \hline
 \hline
 \end{tabular}
 \end{center}
 \caption{Symmetric dissociation of the water molecule, cc-pVDZ basis set.
   The FCI total energy\cite{h2o_fci} is given in $E_{\rm h}$, 
 and the deviations to this reference are given in m$E_{\rm h}$. 
 Comparison with Mukherjee's state specific MRCC values ($E_{\rm Mk} - E_{\rm FCI}$) obtained from Ref.\cite{H2O}. }
 \label{tabh2o}
\end{table}

\section{Properties}
\subsection{Internal decontraction}

The method is internally decontracted. The coefficients of the references as well as those of the Singles and Doubles change along the iterations. 
If the reference space is a valence CAS, treating the non-dynamical correlation
effects, the method takes care of the impact of the dynamical correlation on
the non-dynamical part. The phenomenon is especially important in magnetic
systems where the dynamical charge polarization effects increase dramatically
the weight of the ionic Valence Bond components, diminishing severely the
effective energy of these components.\cite{genesis} This effect is already present in the CASSDCI calculation but the MRCCSD treatment eliminates the size 
consistency defect and slightly improves the quality of the projection of the wave function on the CAS. 

\subsection{Size consistence}
The method does not introduce any unlinked diagram, and is therefore size-consistent. A proof of strict separability 
has been given in the original presentation of the method\cite{Meller_Malrieu_Caballol_1996}. 
It requires that in the splitting into two subsystems 
A and B the active and inactive MOs are localized on one of the two subsystems A or B. 
Actually, as occurs for the Mk-MRCC formalism, the method is not invariant with respect to the unitary transformation of the MOs in their class 
(inactive occupied, active,  inactive virtual). This dependence will be studied in a future work, but the error to FCI being 
small we do not expect a strong dependence on the MO definition. 
As was shown in the study of bond breakings the asymptotic size-consistency error (which is demonstrated to be zero when localized MOs are used) 
is negligible in a basis of symmetry-adapted MOs.

\subsection{Eigenfunction of ${\hat S}^2$}
The here-proposed method does not provide an eigenfunction of ${\hat S}^2$ as we consider only the determinants that are connected by 
an application of ${\hat H}$ to the determinants belonging to the CAS. This treatment does not include higher excitations which will generate 
the full space associated with a given space part. Along all the performed calculations on singlet states, the order of magnitude of the expectation value of 
${\hat S}^2$ calculated on both the CASSDCI and the projected MRCCSD wave functions never exceeded $10^{-3}$. 
A future work will present a solution working with the same restricted space but providing a strict eigenfunction of ${\hat S}^2$.

\section{Prospects}
The formalism presented here allows us to conceive two main types of extensions for further work. 
The first one concerns the reduction of the computational cost of the method through various approximations, in order to target more realistic 
systems. From a methodological point of view, a refined treatment of the excited states deserves to be considered.

\subsection{Computational cost}
The method is extremely flexible, either on the choice of the reference space and/or of the excitations from it. 
One may partition the Singles and Doubles in terms of excitation classes. For instance the most numerous purely inactive excitations 
(2 holes, 2 particles) can be treated in a contracted manner, leading to a diagonal shift of the CASSDCI. 
Another possibility consists in omitting this class of excitation which does not contribute significantly to the vertical energy differences, 
as exploited in the DDCI framework\cite{ddci}. Then, one may exponentialize the semi active excitations and make the DDCI method size consistent. 
As the theory is determinant based, one can take advantage of this flexibility to realize a CIPSI like selection 
of the dominant contributions of both the references and the single and double excitations. 
Further works will investigate the various possibilities such as the combination of MRCC with perturbation theory.  

\subsection{Excited states}

The method is applicable to excited states using several approaches. The
formalism being state specific, the dressing technique of the CI matrix can
be applied to any state dominated by the reference determinants, 
as long as a state following procedure is applied. For states belonging to the same symmetry, 
the resulting eigenvectors will not be strictly orthogonal but might be orthogonalized {\it a posteriori}. 
Another possibility consists in a state average procedure where the amplitudes
are obtained from the values of the quantities $\lambda_i$ averaged over all
desired eigenstates:
\begin{equation}
  \lambda_i = \frac{\sum_m \lambda_i^m (c_i^m)^2 }{\sum_m \left( c_i^m \right)^2 }
\end{equation}
If one refers to the perturbative expression of the first order coefficients,
\begin{equation}
  \lambda_i^m \approx \frac{1}{E_0^m - \langle i | {\hat H} | i \rangle}
\end{equation}
this approximation should be relevant when the states are close enough in energy.

A recent paper\cite{ms_state} has proposed a generalization of this approach to the simultaneous treatment of several states of the same symmetry. 
The basic ideas are the same, except for the fact that the extraction of the amplitudes is more complex. 
The method requires to partition the reference space into a main and an intermediate model spaces, 
in the spirit of the intermediate Hamiltonian formalism. This proposal will be tested in a further work.

\section{Conclusion}

This work shows the relevance of a solution previously proposed to the problem
of the multiple parentage faced by all Multi-Reference treatments, as soon as
the number of targeted vectors is lower than the number of References. The
proposed MRCC algorithm is simple. It only introduces two-body excitation
operators and the number of amplitudes to be determined is reduced to the very
minimum. It proceeds through an iterative dressing of the MRCI matrix,
formulated in terms of a standard eigenvalue equation. It is parallelizable in
the most expensive step (the generation of the coefficients of the Triples and
Quadruples). It is entirely decontracted and may be applied to excited states.
For the list of benchmark studies we have performed, the results are extremely
encouraging.  The present version is state-specific but the principles of
extension to a multi-root version have been formulated. This work actually
opens into several directions, which have to be explored in the future.
The reduction of the computational cost might be done using several
approximations involving the selection of the references and/or Singles and
Doubles according to various criteria. Furthermore, the excited states can be
treated using different approaches, all of them being compatible with the
here-proposed formalism. 

On a different perspective, the multiple parentage problem, which was faced
here in the purpose of building a logically consistent computational tool to go
in the direction of the exact solution, also concerns the building of rational
valence-only effective Hamiltonians. In such an approach, the idea is to map the
information coming from a sophisticated treatment into a minimal effective
Hamiltonian, the parameters of which should be as physically meaningful as
possible.\cite{pradines} We believe that the solution we proposed to the
multiple parentage problem offers a rational solution to this reduction of
information. This remark illustrates the intrinsic link between the two main
tasks of Quantum Chemistry, namely the production of physically grounded
interpretative models on one hand and the conception of rigorous computational
tools.

\bibliography{Publication_MRCC.bib}
\end{document}